\input harvmac
\noblackbox

\overfullrule=0pt
\def\Title#1#2{\rightline{#1}\ifx\answ\bigans\nopagenumbers\pageno0\vskip1in
\else\pageno1\vskip.8in\fi \centerline{\titlefont #2}\vskip .5in}

%
%

\def\a{\alpha}
\def\b{\beta}
\def\c{\gamma}
\def\s{\sigma}

\def\th{\theta}
\def\mod{{\cal M}}
\def\Prop{\noindent {\bf Proposition: }}
\def\Prf{\noindent {\bf Proof: }}
\def\dal{\triangle}
\def\oo{\overline}

\def\L{\Lambda}
\def\delf#1#2{{\partial f^{#1}\over \partial x^{#2}}}
\def\der#1#2{f^{#1}{}_{#2}}
\def\derr#1#2#3{f^{#1}{}_{#2 #3}}
\def\j#1#2{J^{#1}{}_{#2}}
\def\g#1#2{g_{#1 #2}}
\def\om#1#2{\omega_{#1 #2}}
\def\ginv#1#2{g^{#1 #2}}
\def\dt{{d\over dt}}
\def\ds{{d\over ds}}
\def\dft#1{{df^{#1}\over dt}}
\def\dfs#1{{df^{#1}\over ds}}
\def\t#1#2{\theta^{#1}_{#2}}
\def\w#1#2{w_{#1}^{#2}}
\def\tinv{{1\over t}}

\def\p{\partial}

\def\s42{ 2^{-{1\over 4} } }

\def\exp{{\rm exp}}

\font\cmss=cmss10 \font\cmsss=cmss10 at 7pt
\def\IZ{\relax\ifmmode\mathchoice
   {\hbox{\cmss Z\kern-.4em Z}}{\hbox{\cmss Z\kern-.4em Z}}
   {\lower.9pt\hbox{\cmsss Z\kern-.4em Z}}
   {\lower1.2pt\hbox{\cmsss Z\kern-.4em Z}}\else{\cmss Z\kern-.4emZ}\fi}

\def\half{{1\over2}}

\def\p{\partial}
\def\a{\alpha}
\def\b{\beta}

\def\[{\left [}
\def\]{\right ]}
\def\({\left (}
\def\){\right )}
\def\M{{\cal M}}
\def\gab{g_{ab}}
\def\gij{g_{ij}}
\def\gai{g_{ai}}

\def\s{{\sigma}}

\def\app#1#2{\global\meqno=1\global\subsecno=0\xdef\secsym{\hbox{#1.}}
\bigbreak\bigskip\noindent{\bf Appendix:  #2}\message{(#1. #2)}
\writetoca{Appendix {#1.} {#2}}\par\nobreak\medskip\nobreak}
\def\eqnn#1{\xdef #1{(\secsym\the\meqno)}\writedef{#1\leftbracket#1}
\global\advance\meqno by1\wrlabeL#1}
%
%
\def\eqna#1{\xdef #1##1{\hbox{$(\secsym\the\meqno##1)$}}
\writedef{#1\numbersign1\leftbracket#1{\numbersign1}}
\global\advance\meqno by1\wrlabeL{#1$\{\}$}}
\def\eqn#1#2{\xdef #1{(\secsym\the\meqno)}\writedef{#1\leftbracket#1}
\global\advance\meqno by1$$#2\eqno#1\eqlabeL#1$$}

%
\lref\smd{C. Schmidhuber, hep-th/9601003.}
\lref\klm{A. Klemm, W. Lerche, and P. Mayr,
``K3 Fibrations and Heterotic-Type II String
Duality,'' Phys. Lett. {\bf B357} (1995) 313, hep-th/9506112.}
\lref\asplou{P. Aspinwall and J. Louis, ``On the Ubiquity of
K3 Fibrations in String Duality,'' Phys. Lett. {\bf B369} (1996) 233,
hep-th/9510234.}
\lref\leig{ R. G. Leigh, ``Dirac-Born-Infeld Action from
Dirichlet Sigma Model,'' Mod. Phys. Lett. {\bf A4} (1989) 2767.}
\lref\wittf{E. Witten, ``Phase Transitions in M-Theory and F-Theory,''
hep-th/9603150.}
\lref\jpas{J. Polchinski and A. Strominger, ``New Vacua
for Type II String Theory,'' hep-th/9510227.}
\lref\bb{K. Becker and M. Becker, ``M-Theory on Eight-Manifolds,''
hep-th/9605053.}
\lref\hls{J. Harvey, D. Lowe and A . Strominger, ``N=1
String Duality,'' Phys. Lett. {\bf B362} (1995) 65, hep-th/9507168.}
\lref\hala{F. R. Harvey and H. B. Lawson, ``Calibrated Geometries,''
Acta Math. {\bf 148} (1982) 47.}
\lref\bbs{K. Becker, M. Becker, and A. Strominger, ``Fivebranes, Membranes,
and Non-Perturbative String Theory," Nucl. Phys. {\bf B456} (1995)
130, hep-th/9507158.}
\lref\gr{A. Giveon and M. Rocek, ``Introduction to Duality,''
hep-th/9406178, to appear in ``Essays on Mirror Manifolds II,''
International Press.}
\lref\mirr{A collection of introductory articles can be found in
``Essays on Mirror Manifolds,'' S.-T. Yau, ed.,
International Press, 1992.}
\lref\asop{A. Strominger, ``Open P-Branes,'' hep-th/9512059.}
\lref\twn{P. Townsend, ``D-branes from M-branes,'' hep-th/9512062.}
\lref\ks{K. Smoczyk, ``A Canonical Way to Deform a Lagrangian
Submanifold,'' dg-ga/9605005.}
\lref\yz{S.-T. Yau and E. Zaslow, ``BPS States,
String Duality, and Nodal Curves on K3," hep-th/9512121,
to appear in Nuclear Physics B.}
\lref\gsvy{B. Greene, A. Shapere, C. Vafa and S.-T. Yau, ``Stringy
Cosmic Strings and Noncompact Calabi-Yau Manifolds,''
Nucl. Phys. {\bf B337} (1990) 1.}
\lref\cdgp{P. Candelas, X. C. De La Ossa, P. Green, and L Parkes,
``A Pair of Calabi-Yau Manifolds as an Exactly Soluble Superconformal
Theory,'' Nucl.
Phys. {\bf B359} (1991) 21.}
\lref\mcl{R. McLean, ``Deformations of Calibrated Submanifolds,"
Duke preprint 96-01:  www.math.duke.edu/preprints/1996.html.}
\lref\rh{F. R. Harvey, {\sl Spinors and Calibrations,} Academic Press,
New York, 1990.}
\lref\ascon{A. Strominger, ``Massless Black Holes and Conifolds
in String Theory,'' Nucl. Phys. {\bf B451} (1995) 96, hep-th/9504090.}
\lref\amplb{P. Aspinwall and D. Morrison,
``U-Duality and Integral Structures,''
Phys. Lett. {\bf B355} (1995) 141, hep-th/9505025.}
\lref\dmms{D. Morrison, ``Mirror Symmetry and the Type II String,''
hep-th/9512016.}
\lref\bach{B. Acharya, ``N=1 Heterotic/M-Theory Duality and
Joyce Manifolds,'' hep-th/9603033.}

%
\Title{\vbox{\baselineskip12pt\hbox{hep-th/9606040}}}
{\vbox{
\centerline{Mirror Symmetry is T-Duality}}}
\centerline{Andrew Strominger\footnote{$^\dagger$}{email:
andy@denali.physics.ucsb.edu}}
\vskip 0.1in
\centerline{\it Department of Physics, University of California,
Santa Barbara, CA 93106, USA}
\vskip 0.2in
\centerline{Shing-Tung
Yau\footnote{$^{\dagger\dagger}$}{email:  yau@abel.math.harvard.edu}
and Eric Zaslow\footnote{$^{\dagger\dagger\dagger}$}{email:
zaslow@abel.math.harvard.edu}}
\vskip 0.1in
\centerline{\it Department of Mathematics, Harvard University,
Cambridge, MA 02138, USA}

\vskip 0.3 in

\centerline{\bf Abstract}
\vskip 0.1in
It is argued that every Calabi-Yau manifold $X$ with a mirror $Y$
admits a family of supersymmetric toroidal 3-cycles. Moreover
the moduli space of such cycles together with their flat connections
is precisely the space $Y$.  The mirror transformation is
equivalent to T-duality on the 3-cycles.  The geometry of moduli space
is addressed in a general framework.  Several examples are discussed.

\Date{}

\newsec{Introduction}

The discovery of mirror symmetry in string theory \mirr\ has led to
a number of mathematical surprises. Most investigations have
focused on the implications of mirror symmetry of
the geometry of Calabi-Yau moduli spaces. In this paper we
shall consider the implications of mirror symmetry of the
spectrum of BPS soliton states, which are associated to
minimal cycles in the Calabi-Yau. New surprises will be found.

The basic idea we will investigate is briefly as follows.
Consider IIA string theory compactified on a large Calabi-Yau space $X$.
In four dimensions there are BPS states arising from the reduction of
the ten-dimensional 0-brane. The moduli space of this 0-brane is $X$
itself. In four dimensions the 0-brane can be described by a
supersymmetric worldbrane sigma model with target $X$.
The BPS states are the
cohomology classes on $X$, and interactions will involve other invariants
associated to $X$.

Quantum mirror symmetry\foot{This goes beyond
the usual assertion that
the conformal field theories are equivalent, and requires that the
full quantum string theories are equivalent \ascon \amplb \bbs \dmms.} implies
that an
identical
theory is obtained by
compactifying the IIB theory on the mirror $Y$ of $X$. In this
formulation
of the theory, all BPS states arise from supersymmetric 3-branes wrapping
3-cycles in $Y$. (The condition for
supersymmetry is \bbs\ that the 3-cycle is a
special lagrangian submanifold \hala\ and the $U(1)$ connection on
the 3-cycle is flat.)
Hence there must be such a 3-brane in $Y$ whose moduli space is
$X$.\foot{Equality of the BPS spectrum only implies that the
cohomologies are the same, but equality of the
full theory, as well as their perturbation expansions, with all interactions
will require that the spaces are
actually the same. For example if $X$
is large we can localize the 0-brane near a point in $X$ at
a small cost in energy. For very large $X$ the space of such
configurations approaches $X$. Such strong relations are not
possible {\it e.g.} in the context of heterotic-IIA duality
because the perturbation expansions are not directly related.}
The 3-brane moduli space arises both from deformations of
the 3-cycle within $Y$ as well
as the flat $U(1)$ connection on the 3-cycle.\foot{As discussed in
section two, a careful definition of the geometry involves
open string instantons.}
Both of these are generated by harmonic 1-forms on the 3-cycle \mcl,
and the real dimension of the moduli space is accordingly
$2b_1$. Since this
moduli space is $X$ this 3-brane must
have $b_1=3$ in order for the dimensions to match.\foot{It should also
have no self-intersections because massless hypermultiplets at the
intersection point have no apparent analog on the 0-brane side.}
Now the moduli space of the flat connections on a 3-cycle with
$b_1=3$ at a fixed
location in
in $Y$ is a three-torus which, as  we shall see in the following,
is itself a supersymmetric 3-cycle in $X$. Hence this construction
describes $X$ as a three-parameter family of supersymmetric three-tori.
We will refer to this as a supersymmetric $T^3$ fibration of $X$, which in
general has singular fibers (whose nature is not yet well-understood,
but is constrained by supersymmetry).
Consideration of IIB rather than IIA on $X$ similarly
yields a description
of $Y$ as a family of supersymmetric three-tori. Hence the
3-cycles with $b_1=3$ must be three-tori.

To summarize so far,
mirror symmetry of the BPS states implies that every Calabi-Yau $X$ which has
a mirror has a supersymmetric $T^3$ fibration.\foot{This is reminiscent
of the situation found for other types of dualities. For example it
is suspected that a type II compactification has a dual heterotic
representation if and only if the Calabi-Yau admits a $K3$
fibration \klm \asplou.}
The moduli space of the supersymmetric 3-tori together with their
flat connections is then the mirror space $Y$.  Note that this is
an intrinsic formulation of the mirror space.

Now consider the action of T-duality on the
supersymmetric $T^3$ fibers. A 0-brane sitting on
the fiber will turn into a 3-brane, while a 3-brane will turn into a
0-brane. T-duality does not change the moduli space of the
D-brane so this is the same 0-brane - 3-brane pair discussed
above.  We conclude that mirror symmetry is nothing but T-duality on the
$T^3$ fibers!

We will provide a check on this last statement in a certain limit in
section two by an explicit local computation of both the
T-dual and moduli space geometries. The equality of these two
geometries will be seen to follow from the fact that supersymmetric
3-cycles are volume-minimizing. We also discuss the
interesting role
of open-string disc instantons in correcting the moduli
space geometry.

The preceding arguments {\it assume} quantum mirror symmetry.
It is important to understand how much of this structure can
be directly derived without this assumption. As a step in
this direction in section
three we show directly that the moduli space is Kahler and
that it is endowed with a holomorphic $b_1$ form, in
agreement with conclusions following from mirror symmetry.
Section four contains some examples.
Concluding remarks, including a prescription for
constructing M-theory duals for a large class of $N=1$ heterotic
string compactifications, are in section five.

\newsec{Mirror Symmetry and T-Duality}

Consider a family of toroidal, supersymmetric 3-branes $L$ in a Calabi-Yau
space $X$ endowed with metric $g_{MN}$ and 2-form
$B_{MN}$. These are defined by the map
\eqn\map{X^M(\s^i,t^a)}
where $\s^i,~i=1,2,3$ is a periodic spatial coordinate on $L$ and
$t^a, ~~a=1,2,3$ is
a coordinate on the moduli space of supersymmetric maps.
There is also a $U(1)$ connection $A$ on the 3-brane:
\eqn\fcl{A=a_i d\s^i}
Supersymmetry requires \bbs\ that (i) the connection $A$ is flat
(ii) the pullback of the Kahler form on to the 3-cycle vanishes
(iii) the pullback of the holomorphic 3-form on to the 3-cycle
is a constant times the volume element. These last two conditions
imply that the 3-cycle is a special Lagrangian submanifold \hala (cf. section
four also).

We are interested in the full moduli space $\M$, parametrized by
$(a,t)$, of
supersymmetric 3-branes with flat connections.
We will find it convenient to choose local coordinates
\eqn\lcl{X^i=\s^i, ~~X^{a+3}=t^a,~~~a=1,2,3.}
The line element on $X$ is then
\eqn\lmt{ds_X^2=\gab dt^adt^b +2\gai dt^ad\s^i +\gij d\s^id\s^j.}
\lmt\ locally describes a $T^3$ fibration of $X$.
The two form is
\eqn\bjj{B=\half B_{ab} dt^a \wedge dt^b +B_{ai} dt^a\wedge d\s^i +
\half B_{ij} d\s^i\wedge d\s^j .}
In general there will of course be
singular points at which the fibers degenerate.

We wish to compute the metric on $\M$. We first consider the
tree-level contribution. This is derived
from the Born-Infeld action for the 3-brane
\refs{\leig, \twn}
\eqn\trb{S_3 =-\int d^4\s \sqrt{-det(E-F).}}
where
\eqn\sdf{\eqalign{F_{\mu\nu}&=\p_\mu A_\nu-\p_\nu A_\mu,\cr
E_{\mu\nu}&=E_{MN}\p_\mu X^M
\p_\nu X^N,\cr E_{MN}&=g_{MN}+B_{MN}, \cr \mu,\nu,\a,\b&=0,1,2,3,\cr}}
and we have fixed the string coupling.

To describe slowly time-varying configurations we insert the ansatze
\eqn\ans{\eqalign{ \dot X^{a+3}&=\dot t^a, \cr \dot X^{i}&=-E^{ji}E_{aj} \dot
t^a, \cr
F_{0i}&=\dot a_i,\cr}}
where the dot denotes differentiation with respect to time $\s^0$, and
$E^{ij}E_{jk}=\delta^i{}_k$.\foot{Note that $E^{ij}$ is the inverse
of the tensor $E_{ij}$ on $L.$  So, for example, $E^{ia}$ makes no
sense, and $E^{ij}E_{ja} \neq 0$ in general.}
The $\dot X^i$ term is arranged so that the motion of the 3-brane is normal to
itself, with a $B$-dependent rotation in $L$.
Expanding \trb\ to quadratic order in time derivatives, the leading
kinetic part of the action is then
\eqn\trb{\eqalign{S_3&=\int d^4\s \sqrt{-E}[ (E_{ab} -E_{ai}E^{ij}E_{jb})
\dot t^a \dot t^b \cr &\phantom{int}+(E^{ij}E_{ja}-E_{aj}E^{ji})\dot a_i
\dot t^a +E^{ij}\dot a_i\dot a_j
)] +....\cr}}
Hence the tree-level line element on $\M$ is
\eqn\mlm{ds_\M^2=(E_{ab} -E_{ai}E^{ij}E_{jb})
d t^a d t^b +(E^{ij}E_{ja}-E_{aj}E^{ji})d a_i
d t^a +E^{ij}d a_i d a_j.}

Mirror symmetry implies that the full metric on $\M$ equals the metric on
the mirror $Y$ of $X$. The leading result
\mlm\ can not be the full story in general.
The metric \mlm\ has a $U(1)^3$ isometry generated by shifts in $a_i$.
This is certainly not present for a generic Calabi-Yau, so \mlm\ can not
in general be the exact metric on the mirror. However there
are instanton corrections which can resolve this apparent discrepancy.
The motion of the 3-brane is generated
by open strings with Dirichlet boundary conditions on the 3-brane.
The metric on $\M$ is defined by a two point function on a
disc whose boundary is the 3-brane. This receives instanton corrections
from minimal area discs whose boundaries wrap a nontrivial 1-cycle in $L$.
(If $Y$ is simply connected all such cycles bound discs.) Generically
such corrections become exponentially small at large radius. However
even at large radius the corrections are non-negligible near the singularities
of the
fibration where the $T^3$ degenerates. After including these corrections
mirror symmetry predicts that \mlm\ will agree with the metric on $Y$.
This is quite analogous to the usual mirror story where the flat
metric on the complexified Kahler cone agrees with the metric
on the space of complex structures only after including world sheet
instanton
corrections.

Instanton corrections to the moduli space geometry of 3-cycles in
$X$ are suppressed
in the limit of large radius away from the singular
fibers. The moduli space geometry is then the
large complex structure limit of $Y$. In this limit
the metric
takes the symmetric form \mlm\ and hence admits a local $U(1)^3$ action away
from the singular fibers.
We refer to this as a ``semi-flat" Calabi-Yau metric.
A lower-dimensional example of such a metric is found in
equation 4.1 of \gsvy, where stringy cosmic strings are described as
a flat $T^2$ fibered over $R^2$ with a singularity at the origin.\foot{The
resolution
described in \gsvy\
of those singularities can be understood as arising from disc
instanton corrections.}  A semi-flat $K3$ metric can be obtained by patching
together 24
such singular cosmic strings. A 6-dimensional example can be obtained by simply
tensoring this $K3$ with $T^2$. The semi-flat metric on the quintic
will be discussed in section four.

Because there is locally a $U(1)^3$ symmetry, an equivalent string theory can
be explicitly constructed locally (away from the singularities)
by T-duality on the flat $T^3$ fibers. This duality
interchanges IIA and IIB, and maps
IIA 0-branes and supersymmetric IIB 3-branes tangent to the fibers
into one another. Hence this T-duality is the same as mirror symmetry.

It is illuminating to check in detail by a local
calculation that the T-dual and moduli space geometries are
indeed the same.
The string sigma model action on a flat worldsheet is
\eqn\ssm{S_\s={1 \over 2 \pi}\int (E_{ab}\p t^a \bar \p t^b +E_{ia}\p \s^i
\bar \p t^a + E_{ai}\p t^a\bar\p \s^i +E_{ij} \p \s^i \bar \p \s^j ). }
The three symmetries of shifts of the $\s^i$ imply that the
three currents
\eqn\curr{j_i=E_{ji} \p \s^j +E_{ai} \p t^a }
are conserved. We may therefore define three real worldsheet
scalars $a_i$ by
\eqn\adf{j_i=i\p a_i.}
Eliminating the scalars $\s^i$ in favor of $a_i$ one finds an equivalent T-dual
sigma model. The T-dual metric is precisely \mlm, see for example \gr\
where various subtleties in the transformation are also discussed.

In general the T-dual of a Ricci-flat metric obeys the low energy
string equations of motion. However it is not necessarily itself
Ricci-flat because T-duality generically generates an
axion field strength $H=dB$ and a dilaton field
$\Phi$ with a nonvanishing gradient, both of which
 act as a source for the Ricci tensor.
For simplicity consider in the following the special case in which
$B$ is zero before T-dualizing. The new $\Phi$ is then
\eqn\phii{\Phi={\rm ln ~det}g_{ij}.}
The new $H$ is
\eqn\hng{H^i_{~~ab}={g^{ij}}_{,a}g_{jb}+g^{ij}g_{jb,a}-
{g^{ij}}_{,b}g_{ja}-g^{ij}g_{ja,b}.}
The metric $g$ on $X$ in the coordinates (2.4) obeys various
differential identities by virtue of the fact that
all three-tori at constant $t^a$ minimize the volume \bbs.
These constraints on $g$ imply that
$H$ vanishes and $\Phi$ is constant after T-duality, as required.

To see this consider a small normal perturbation of the three-torus at $t=0$
described by
\eqn\prt{\eqalign{X^i &=\s^i-g^{ij}(0)g_{ja}(0)\L^a(\s), \cr X^a &=
\L^a(\s).\cr}}
Through first order in $\L$ and first order in $\p \L$ the induced metric
$h$ is given by, in the semi-flat case where the metric is
independent of $\s$,
\eqn\hind{\eqalign{h_{ij}&=\bigr[g_{ij}+g_{ij,a}
\L^a
+g_{ia,b}
\L^b\p_j\L^a+g_{ja,b}\L^b\p_i \L^a\cr&~~-g_{il,b}g^{lk}g_{ka}\L^b\p_j\L^a-
g_{jl,b}g^{lk}g_{ka}\L^b\p_i\L^a
\bigl]_{t=0}.}}
The change in the volume may be written, using \phii\ and \hng\ and
integrating
by parts
\eqn\vch{\eqalign{\delta V_3&=\int d^3\s \delta \sqrt{h} \cr
&=\half \int d^3\s \sqrt{g} \bigl( \p_a\Phi\L^a
 +H^i_{~~ab}\L^a\p_i\L^b
\bigr).}}
The fact that the area does not change at first order in
$\L$ implies that the dilaton is constant.
The $\L\p\L$ variation
will be non-negative for all $\L$ if and only if $H=0$.
Hence $H$ vanishes and the T-dual metric must be Ricci-flat.

At this point it is easy to see that, as mentioned in the introduction,
the three-tori defined by $t=constant$ in the moduli space metric
\mlm\ are themselves supersymmetric
3-cycles in the mirror $Y$ of $X$. One way of stating the condition
for supersymmetry is that the T-dual geometry should have
vanishing $H$ and constant $\Phi$. This is obviously the case
since T-dualizing just returns us back to $X$.

Now consider perturbing the Calabi-Yau moduli so that the metric is no longer
semi-flat. In that case the action of T-duality on the fibers still exists
but is complicated because there
are no isometries. The construction of the moduli space $\M$ of supersymmetric
3-branes is complicated by instantons. Nevertheless since they are both
described as perturbations of the same theory the equivalence between
$T^3$ T-duality and mirror symmetry should be valid for all Calabi-Yau
geometries in a neighborhood of the semi-flat geometries.
Furthermore
since the supersymmetric
3-cycles correspond to minimally-charged BPS states we expect
that they survive sufficiently small perturbations. In conclusion the
relation between fiberwise T-duality and mirror symmetry is
quite general and not restricted to the semi-flat limit.

\newsec{D-brane moduli space}

We now turn to some mathematical generalities of the above discussion.
We will not be able to prove all the statements of the previous
section regarding D-brane moduli
space, but we hope to provide a framework for future work on these
matters.  We will show that with a certain natural metric, the D-brane moduli
space has a K\"ahler structure.  In addition, there are several
natural forms on this moduli space which suggest that its interpretation
as a Calabi-Yau for toroidal D-branes makes good sense.
In fact, a ``dual" geometry emerges which corresponds to the
mirror symmetry interpretation of the previous section.

Recall that we are interested in a special Lagrangian submanifold, $L,$
of a Calabi-Yau manifold, $M.$  Let $f: L \rightarrow M$ be the map
of the imbedding (generally, $f$ will only need to be an immersion).
The special Lagrangian condition is equivalent (\rh, Corollary 7.39)
to the statement that
$$f^*\omega = 0 \qquad \hbox{and} \qquad f^*\kappa = 0,$$
where $\omega$ is the K\"ahler form and $\kappa$ is
the imaginary part of the Calabi-Yau form on $M.$  The asterisk
denotes the pull-back operation.  We will study the space of
special Lagrangian immersions $f$ (which arise as deformations
of a fixed immersion) and call this space ${\cal M}_{sl}.$

We use the following conventions:

$\bullet\quad$ $\alpha, \beta, \gamma, ...$ are indices for $M.$

$\bullet\quad$ $y^\alpha, y^\beta, ...$ are coordinates on $M.$\foot{We will
not use complex
coordinates for $M;$ we will use real coordinates adapted to
the complex structure, which will then
be constant in these coordinates.}

$\bullet\quad$ $i,j,k, ...$ are indices for $L.$

$\bullet\quad$ $x^i, x^j, ...$ are coordinates on $L.$

$\bullet\quad$ $\mod_{sl}$ is the moduli space of special Lagrangian
submanifolds.

$\bullet\quad$ $a, b, c, ...$ are indices on $\mod_{sl}$

$\bullet\quad$ $t_a, t_b, ...$ are coordinates on $\mod_{sl}.$

$\bullet\quad$ $g_{\alpha\beta}, g_{ij}, g_{ab}$ are metrics on
$M, L, \mod_{sl}.$\foot{This notational
redundancy begs the question, ``What is $g_{12}$?"  We
will be careful never to use numbers where our indices
stand, and operate under the assumption that an
additional identifier would be employed (e.g., $\oo{g}_{12}$ on $M$).}
Without indices, we write $\oo g, g, g_{SL}.$

$\bullet\quad$ $J^\alpha{}_\beta$ is the complex structure on $M.$  $\j \a \c
\j \c \b =
- \delta^\alpha{}_\beta.$

$\bullet\quad$ $\omega_{\alpha\beta} = g_{\alpha\gamma}J^\gamma{}_\beta.$

$\bullet\quad$ $\der \a i = \delf \alpha i,$ etc.

$\bullet\quad$ All indices are raised or lowered by the appropriate metric.

$\bullet\quad$ For a Lagrangian submanifold, a tangent
vector $w^ie_i$ (where
$e_i = {\del\over \del x^i}$) determines
a normal vector $\underline{w} = w^i(J\cdot f_* e_i),$ and its associated
second
fundamental form $(h^{\underline{w}})_{jk}$ is given by
$w^i h_{ijk},$ with $h$ a symmetric tensor defined by
$h_{ijk} = - \langle
\oo{\nabla}_{f_* e_i} f_* e_j, J \cdot f_*e_k \rangle$ \ks.

As yet, we have not defined the coordinates on moduli space, or
even shown that it exists as a manifold.\foot{This
was shown on general grounds in \mcl, though without the
introduction of coordinates, which we shall need for our
geometric analysis.}  We do so now,
defining coordinates analogous to Riemann normal
coordinates, which give a coordinatization of a Riemannian
space by the tangent space of a point.  Our point will be
the initial special Lagrangian submanifold, $L,$ with
the map $f.$

The tangent space to $\mod_{sl}$ at $L$ can be found by
considering an arbitrary one-parameter deformation $f(t)$ of
$f = f(0).$  In order to remove ambiguity, we require throughout
that $\dot{f},$ a vector field on $f(L)$ in $M,$ be a normal
vector field.  This can be achieved through a $t$-dependent
diffeomorphism of $L.$  Now $\dot{f} =  {d f^\alpha \over dt}{\del\over\del
y^\alpha}$ is orthogonal to $f(L).$  Therefore, $J\cdot \dot{f}$ is a
tangent vector (by the Lagrangian condition), which we can convert
to a 1-form by the metric $\oo{g}$ on $M.$
This combination gives us a 1-form ${d f^\beta\over dt} \omega_{\beta\alpha}
dy^\alpha.$  We define $\theta$ to be the pull-back of this 1-form to
$L;$ $\;\theta_i = \dot{f}^\b \om \b \a \der \a i.$\foot{Note
that the definition of $\theta$ is independent
of the requirement that $\dot{f}$ be normal, for any tangent vector
added to $\dot{f}$ adds nothing to $\theta,$ by the Lagrangian
condition.}

\Prop ${d\over dt}f^*\omega = d\theta.$

\Prf Commuting derivatives and using the closure of $\omega,$ one computes
$$\eqalign{{d\over dt}&f(t)^*\omega_{ij}(x) = {d\over dt}
\left(\der \a i \der \b j \om \a \b (f(x))\right)\cr
&= \del_i(\dot{f}^\a ) \der \b j \omega_{\a\b} + \der \a i
\del_j (\dot{f}^\b )\om \a \b + \der \a i \der \b j \dot{f}^\c \del_\c
\om \a \b \cr
&=
\del_i(\dot{f}^\a ) \der \b j \omega_{\a\b} + \der \a i
\del_j (\dot{f}^\b )\om \a \b +  \der \a i \der \b j \dot{f}^\c (-\del_\a
\om \b \c - \del_\b \om \c \a ) \cr
&=
\del_i(\dot{f}^\a ) \der \b j \omega_{\a\b} + \der \a i
\del_j (\dot{f}^\b )\om \a \b + \( \der \b j \dot{f}^\c \del_i \om \c \b
- \der \a i \dot{f}^\c \del_j \om \c \a \) \cr
&=
\del_i\( \dot{f}^\a \om \a \b \der \b j \)
- \del_j \( \dot{f}^\b \om \b \a \der \a j \) \cr
&= (d\theta)_{ij}.}
$$
Therefore, to preserve the Lagrangian condition (to first order), the normal
vector from the deformation $f(t)$ must give rise to a 1-form which pulls
back to a closed 1-form $\theta$ on $L.$

Likewise, we have the following proposition.

\Prop ${d\over dt}f^*\kappa = 0 \Rightarrow d^\dagger \theta = 0.$

The proof, similar to (though not quite as straightforward as)
the above, may be found in \mcl.
Therefore, $\theta$ must be closed and co-closed, i.e. harmonic.

The tangent space to deformations of isomorphism classes of flat $U(1)$ bundles
is simply described.  If $A$ represents a flat $U(1)$ connection on a bundle
$E\rightarrow L,$ then a deformation of the connection has the form $A +
\theta.$  Requiring this to be flat means that $d(A+\theta)=0$ and hence
$d\theta = 0,$ since $A$ is flat.  Further, $A$ (equivalently $\theta$) is only
defined up to gauge transformations $A\sim A+d\chi.$  Therefore, the tangent
space to the space of flat bundles is closed 1-forms modulo exact 1-forms,
i.e. $H^1(L).$  A Riemannian metric on $L$ -- in our case the pull-back
metric --
provides an isomorphism $H(1)\cong {\cal H}^1(L).$  In fact the full space of
flat $U(1)$ bundles is described by the set of monodromies, $\exp{[2\pi i
c_j]},$ $j = 1...b_1(L),$ where $b_1$ is the first Betti number.
This space is the torus $T^{b_1}.$  The moduli space $\mod$ is
a fibration over the moduli space of special Lagrangian manifolds
(continuously connected to $f(0)(L)$), with fiber $T^{b_1}.$

Therefore, the tangent space to $\mod$ at $L$ is ${\cal H}^1(L)\oplus {\cal
H}^1(L),$
and there is a natural metric on ${\cal H}^1,$ which defines for us a
metric on $\mod_{sl}$ and a block diagonal metric on $\mod.$
Let $\theta^a$ and $\theta^b$ be two harmonic 1-forms on $L,$ with
its induced metric $g_{ij}.$  Then
$$g_{\cal M} = \pmatrix{g_{SL}&0\cr 0&g_{SL}},$$
where
$$(g_{SL})_{ab} = \int_L \theta^a_i\theta^b_j g^{ij} \sqrt{g} dx.$$
There is also a natural almost complex structure, which takes the form
$${\cal J} = \pmatrix{0&{\bf 1}\cr -{\bf 1}&0}.$$
The coefficients of this matrix might not be constant in a coordinate frame
for $\mod,$ so we have not yet shown that it is a complex structure, or
that the metric is K\"ahler with respect to it.  We shall do this
presently, after defining a coordinate frame.

Knowing that the tangent space to $\mod$ at $L$ is ${\cal H}^1\oplus {\cal
H}^1$ is not enough for us to be able to set up coordinates for $\mod.$  To do
this, we need something analogous to the exponential map for Riemann normal
coordinates.  In this case, one needs, given a tangent direction, a uniquely
defined one-parameter family of flows of $L$ in $M$ which are special
Lagrangian, i.e. for which $f(t)^*\omega = f(t)^*\kappa = 0$ for all $t.$
Consider a harmonic 1-form $\theta$ on $L,$ with $w$ its corresponding
vector, $w^i = g^{ij}\theta_j.$  $w$ obeys $\nabla_iw^i = 0.$  By continually
pushing forward $w$ and rotating it by the complex structure, we can attempt to
define a flow through the equation
$$\dot{f}(t) = J\cdot f(t)_* w.$$
However, there is no guarantee that the manifolds $f(t)(L)$ will continue to be
special Lagrangian manifolds at all times.  The problem is that $g = f^*\oo{g}$
evolves in $t,$ and so $\theta$ may not remain harmonic.  To remedy this, we
{\sl require} $\theta$ to be harmonic by including a time-dependence
explicitly.  We do so in such a way that $[\theta],$ the cohomology class
defined by $\theta,$ does not have a $t$-dependence.  Therefore, we define
$$\widetilde{\theta}(t) = \theta - d\phi(t);$$
then the requirement that $\widetilde{\theta}$ be harmonic becomes
$\nabla_i (g^{ij}\del_j\phi) = \nabla_i w^i$
or
$$\triangle \phi = \nabla_i w^i = d^\dagger \theta.$$
Do not forget the implicit $t$-dependence of the operators!  Note that at a
given time $t,$ the last equation determines $\phi$ up to an additive constant,
which is irrelevant to the definition of $\widetilde{\theta}(t).$

We can now try to define a one-parameter flow starting from $L$ with initial
normal vector $J\cdot f(0)_*w$ as the solution to the coupled equations
\eqn\slflow{\eqalign{{df^\alpha\over dt} &= J^\alpha{}_\beta{\del f^\beta\over
\del x^i}g^{ij}(\theta{}_j - \del_j\phi)\cr
\dal\phi &= d^\dagger \theta,}}
where $\theta$ has no $t$-dependence.
This is analogous to geodesic flow on Riemannian manifolds (though we don't
claim that this flow is geodesic with respect to the metric on moduli space).

\Prop This flow exists, at least infinitesimally, and is unique.

\Prf A proper demonstration of the above uses results we will
derive in this section.  Basically, one can recursively solve for every
derivative of $f$ with respect to $t.$  At first order, the bottom equation
specifies $\phi = {1\over \triangle}d^\dagger \phi + constant,$
which gives a unique expression for the top equation (the
constant falls out).  To find ${d^2f\over dt^2}$ one can differentiate
the bottom equation to get $\triangle \dot{\phi} = d^\dagger \psi,$
for some $\psi$ (specifically, $\psi_j =
2(h^{\underline{w}})_{jk}g^{kl}(\del_l\phi - \theta_l)$), and so
on for the higher derivatives.  We have not determined the radius
of convergence (possibly zero) for this expansion in $t.$  For the
remainder of this paper, we assume that this flow exists for finite time.

Let $\theta^a,$ $a  = 1...b_1$ be a basis for $\cal{H}^1(L),$
$w_a$ the corresponding vectors, and $t_a$ the corresponding
coordinates in moduli space.  We call $L_t$ the submanifold
$f(1)(L),$ where the flow is defined by the 1-form $\theta = t_a\theta^a.$
We can coordinatize the full
moduli space $\mod$ with $t_a$ and $s_a,$ $a = 1...b_1,$
where the $s_a$ simply tell us to use the connection
$A + s_a \theta^a$ on $L_t.$  Now while we know that $\theta$ does
not change in cohomology during the flow it defines, other
1-forms might do so.  As a result, the almost complex structure, ${\cal J},$
could pick up a coordinate dependence.\foot{If $[\theta^b(t)]$
is not constant, write $[\theta^b(t)] = y^b{}_c(t)\theta^c.$
Then ${\cal J} = \pmatrix{0 &y^{-1}\cr -y&0}.$}  The calculation in the
appendix will show that, to first order, this does not occur,
and as a result the Nijenhuis tensor vanishes.  This shows that
${\cal J}$ is indeed a complex structure on moduli space.

We turn now to the geometry of the flow.  Let
$f(0)$ be an imbedding and $f(t)$ a family of special
Lagrangian immersions inducing   Then we have the following result.

\Prop ${d\over dt_a}g_{ij} = 2h_{ijk}w_a^k,$ which is twice
$(h^{\underline{w}})_{ij},$ the
second fundamental form defined by the normal vector $J\cdot w.$

\noindent
Here
$$h_{ijk} = - \langle
\oo{\nabla}_{f_* e_i} f_* e_j, J \cdot f_*e_k \rangle,$$
and $\oo{\nabla}$ represents
the covariant derivative on $M.$  It is not hard to show
that by the Lagrangian property $h$ is a symmetric three-tensor
on $L.$  These and other properties of Lagrangian flow
are investigated in \ks.  The proof of the proposition is straightforward --
one simply puts $g_{ij} = \der \a i \der \b j \g \a \b$ and differentiates,
substituting \slflow\ for time derivatives (note that ${d\over dt_a}\g \a \b
= \dot{f}^\c \del_\c \g \a \b$).

As with Riemann normal coordinates, we have to express the
vector field defined by $\theta^b$ as a function of the coordinate
$t$ (analogous to push-forward by the exponential map).
As a result, the forms pick up a $t$-dependence.  We
hope to find that our flow gives
the $\theta^a$ a $t$-dependence such that the cohomology class
does not change as a function of time, but note that the form itself
must change, since it must remain harmonic with respect to the
changing metric.  That is, we want that ${d\over dt_a}[\theta^b] = 0,$
where the brackets represent cohomology (the flows
are already defined so that ${d\over dt_a}[\theta^a] = 0,$ no sum) --
i.e., that ${d\over dt_a}\theta^b = d\psi^{ab}$
for some function $\psi^{ab}(t;x).$
We cannot solve the flows for finite time, but we can compute
${d\over dt_a}\theta^b\vert_{t=0}$ as follows.  Consider a one-parameter
family of flows $f_r(t)$ defined with initial vector
$\theta^a + r\theta^b.$  These define a two-dimensional cone in moduli
space, and we can take the derivative in the $b$ direction at time
$t_a$ by computing ${1\over t_a}{d\over dr}\vert_{r=0}.$\foot{If we
write $t_a,$ we mean
to set $\vec{t} = (0,...,t_a,...0),$ with lone
nonvanishing $a^{\hbox{th}}$
component.}  This gives a normal
vector (it may not be normal, but the pull-back to a 1-form will
be insensitive to tangent directions), which then defines a 1-form
representing $\theta^b(t_a).$  Specifically,
\eqn\thb{\theta^b(t_a)_i = -{1\over t_a}J^\alpha{}_\beta
\left({d f_r^\beta\over dr}\vert_{r=0}\right)
g_{\alpha\gamma}\delf{\gamma}{i},}
where the family of flows is defined by the equations
$$\eqalign{{df^\alpha \over dt}&= J^\alpha{}_\beta\delf{\beta}{i}g^{ij}
(\theta^a_j + r\theta^b_j - \del_j\phi)\cr
\triangle \phi &= d^\dagger(\theta^a + r\theta^b).}$$
Do not forget the $r$ and $t$ dependence hidden in the second
equation.  Note, too, that $\theta^a$ and $\theta^b$ have no
$t$- or $r$-dependence.

We can see that $\hbox{lim}_{t_a\rightarrow 0}\theta^b(t_a) = \theta^b(0),$
as it must, by L'H\^opital's rule and some algebra, noting that
$f(0)$ is independent of $r.$

It remains to compute $\hbox{lim}_{t_a\rightarrow 0}{d\theta^b(t)\over dt_a}.$
This computation, performed in the appendix, depends on using L'H\^opital's
rule and applying the definition of the flow for each time derivative.
After some algebra, one finds
\eqn\calc{{d\theta^b(t_a)\over dt_a}\vert_{t=0} = -\half d
\left(d^2\phi\over drdt\right)_{r=t=0} \equiv d\psi^{ab}.}

This shows, too, that
\eqn\symg{\eqalign{\del_ag_{bc} &= {d\over dt_a}\int \theta^b(t)_i
\theta^c(t)_j g^{ij}
\sqrt{g}dx \cr
&= \int \langle d\psi^{ab}, \theta^c \rangle +
\int \langle \theta^b,d\psi^{ac} \rangle - 2\int
h_{ijk}w_a^iw_b^jw_c^k\sqrt{g}dx \cr
&= \del_bg_{ac},}}
where we have used harmonicity and the symmetry of $h_{ijk},$
as well as the fact that the derivative of $\sqrt{g}$ is
proportional to the mean curvature, which vanishes for minimal
manifolds ($h$ is traceless).
The result \calc\ also shows
that the derivatives of ${\cal J}$ are zero, which immediately
implies that it is a complex structure.\foot{The
Nijenhuis tensor, ${\cal N}^c{}_{ab} = {\cal J}^d{}_a
(\del_c{\cal J}^c{}_b - \del_b{\cal J}^c{}_d) - ( a \leftrightarrow b ),$
clearly vanishes.}  In addition,
\symg\ tells us that the K\"ahler form on moduli space,
$\omega^\mod_{ab} = g_{ac} {\cal J}^c{}_b,$ is closed at $L,$ which
was an arbitrary point on $\mod.$

On $\mod_{sl},$ there exists a natural $n$-form (where $2n$
is the real dimension of $M,$) $\Theta,$
defined as follows \mcl:
$$\Theta(\theta^{a_1},...,\theta^{a_n}) = \int_L \theta^{a_1}\wedge
...\wedge \theta^{a_n}.$$
Now since in our coordinates the $\theta^{a_i}$ change by exact
terms, all derivatives of $\Theta$ vanish, and $\Theta$ is closed.
Further, we can extend $\Theta$ to a form on $\mod$ by defining,
$$\Theta^\mod = \Theta_{a_1...a_n}dz^{a_1}...dz^{a_n},$$
where $dz^a = dt_a + ids_a.$
The assertion that $d\Theta^\mod = 0$
follows from the above, and fact that there is no $s$-dependence.
We could have complex conjugated
some of the $dz$'s and obtained other forms.  We don't know how
many of these forms will survive the compactification procedure,
but we point out that in the special case of toroidal submanifolds,
$b_1 = n$ and so
$\Theta^\mod$ is a holomorphic $b_1$-form, which could be the
Calabi-Yau form. This lends support, at least, to the conjectures
of the previous section.  Note, too, that one
can also, given a 2-form potential $B_{\a\b},$
define a 2-form on moduli space $B_{ab} = \int B^{ij}\theta^a_i\theta^b_j
\sqrt{g},$ where $B_{ij}$ are the components of $f^*B,$ and
coordinates are raised, as usual, by $g^{ij}.$

Of course, there is more about this moduli space to study.
We need to determine its curvature.  The variation
of the metric means that our coordinates are not normal.  A coordinate
transformation mixing $t$ and $s$ is needed to make it so.  Therefore,
the fibration may not be holomorphic with respect to this metric.
This makes some sense, as we don't expect, for example for K3, to find
the fibers always perpendicular to the base.  However, we do not yet
know how this metric should get corrected, or what the proper compactification
of moduli space looks like.  We do not know how the K\"ahler
class gets corrected, as a result.  We hope more rigorous mathematical
results will ensue.

\newsec{Examples}

\subsec{K3}

The only well-understood example of the preceding is K3 (or $K3\times T^2,$
if we want a three-fold).  The supersymmetric cycles can be related, by
a rotation of complex structures\foot{The K3 has three complex
structures, $I, J, K,$ giving three 2-forms,
$\omega_I, \omega_J, \omega_K$ which can be interchanged by SO(3) rotations.
With complex structure $J,$ we have $\omega = \omega_J$ and
$\Omega_J = \omega_I + i\omega_K,$ so we can relate
the condition $f^*(\omega) = f^*(\hbox{Im}\Omega_J) = 0$ with
complex structure $J$ to $f^*(\Omega_K) = 0,$ in complex structure
$K.$  These surfaces are then holomorphic submanifolds with respect
to $K.$} to Riemann
surfaces sitting in the K3.  For genus one, these are tori, which
are easily apparent if the K3 is written as an elliptic fibration:
$\pi: K3 \rightarrow {\bf P}^1.$
Each point $p$ on the K3 determines a torus $\pi^{-1}(\pi(p)),$
and exists as a point on
that torus (which is its own Jacobian).  Given a section, this uniquely
determines a submanifold and a flat bundle on it.  Thus K3 appears as
the moduli space for submanifolds which are tori.  This is appropriate,
as K3 is its own mirror!

\subsec{The Quintic}

The example of the quintic has been analyzed in great detail \cdgp.
It can be seen as follows that, at least for sufficiently
large complex structure, there is a supersymmetric $T^3$ cycle
in the homology class predicted by mirror symmetry.
On the mirror quintic $Q$  there is a
a 0-brane, with moduli
space equal to $Q$. There is also a 6-brane
which wraps $Q$ and has no moduli (note that there are no flat bundles
since $Q$ is simply connected). Quantum mirror symmetry
tells us that the corresponding 6-
and 0-cycles should be mirror to two distinct 3-cycles on the
quintic, $\widetilde{Q}$. The mirror transformations
relating the even cohomology on $Q$ to the odd cohomology
on $\widetilde{Q}$ have been explicitly displayed in \cdgp.
Using these transformations it was shown in \jpas\ that the
6-cycle on $Q$ is mirror to the 3-cycle on $\widetilde{Q}$
which degenerates at the conifold. Since the 0-cycle is Poincare
dual to the 6-cycle (at large radius where
instanton corrections can be ignored) it must be mirror to a
3-cycle Poincare dual to the one which degenerates at the
conifold.
Consider the 3-cycle defined in \cdgp\ by
\eqn\ttr{|z_1|=a,~~|z_2|=b,~~|z_3|=c,}
where $a,~b,~c$ are real. Consider a patch in which we can fix $z_5=-1$
and take $z_4$ to be a root of the polynomial equation
for a quintic
\eqn\qnt{z_1^5+z_2^5+z_3^5+z_4^5=1.}
If $a,~b,~c$ are sufficiently small the roots are non-degenerate
and as the phases of $z_1, ~z_2, z_3$ vary a three-torus is swept out
in  $\widetilde{Q}$. Topologically \ttr\ defines a
$T^3$ ``fibration'' of $\widetilde Q$ with base parameterized by $a,~b,~c$.
The fibers become singular at values of $a,~b,~c$ for which the
roots collide. It was shown in \cdgp\ that this 3-cycle is
Poincare dual to the vanishing cycle at the conifold. Hence it is in
the correct homology class to be mirror to the 0-brane on $Q$.
To see that there is actually a supersymmetric cycle in this
class consider the large complex structure limit in which
a term $\lambda z_1z_2z_3z_4z_5$ is added and the coefficient
$\lambda$ taken to infinity. There is a branch on which we
may take $z_5=-1$ and $z_4$ near zero and the metric approaches
(up to a constant)
\eqn\metlm{ds^2=d\ln z_1 d \ln \bar z_1 +d\ln z_2 d \ln \bar z_2
+d\ln z_3 d \ln \bar z_3 }
for finite  $z_1, z_2, z_3$.
In this limit \ttr\ obviously defines a family of supersymmetric 3-cycles.
Perturbing away from the large complex structure limit
produces a small effect on the Ricci-flat metric for finite
 $z_1, ~z_2, z_3$. Since supersymmetric 3-cycles are expected to be
stable under
small
perturbations of the metric, such a 3-cycle should exist in a neighborhood
of the large complex structure limit.

\newsec{Conclusions}

Our results should lead to the construction of a large new class of
dual
pairs. For example, heterotic string theory on $T^3$ is
equivalent to M-theory on $K3$. Applying this
duality fiberwise to the
$T^3$ fibers of any $N=1$ compactification of heterotic
string theory on a Calabi-Yau space with a mirror
partner, one obtains an $N=1$ compactification of M-theory
on a seven-manifold of $G_2$ holonomy.
Examples of this type are in \hls \bach.
By considering
perturbations this duality can be extended to $(0,2)$
heterotic compactifications.\foot{Recent progress on this problem has
been made using F-theory duals in \wittf.}
Perhaps this will provide a useful
way to study a phenomenologically interesting class of string
compactifications. A precise understanding of this duality, as
well as the mirror relation discussed herein, will probably require
a better understanding of the nature of the singular fibers.
Undoubtedly, constraints from supersymmetry will play a role in
controlling these singularities.

In this paper we have only considered the
implications of quantum mirror symmetry  for the simplest case
of a single 0-brane. We expect that consideration of other
p-branes and their bound states will lead to further insights
into the rich structure of Calabi-Yau spaces
and supersymmetric string
compactifications in general.

\app{A}{The calculation of ${d\over dt_a}\theta^b(0).$}

We calculate ${d\over dt_a}\theta^b(0),$ where conventions
are as in section three.
Let us now use $t$ for $t_a.$
We have from \thb\ that $\theta^b(t)$ is of the
form $\theta^b(t) = C/t,$ where $C\rightarrow 0$ as
$t\rightarrow 0$ (since ${df_s\over ds} = 0$ at $t=0$).  So
$$\eqalign{\hbox{lim}_{t\rightarrow 0}{d\theta^b\over dt} &=
\hbox{lim}_{t\rightarrow 0}(t\dot{C}-C)/t^2 \cr
&= \hbox{lim}_{t\rightarrow 0}(t\ddot{C} + \dot{C} - \dot{C})/2t
\qquad\hbox{by L'H\^opital's rule}\cr
&= \hbox{lim}_{t\rightarrow 0}\ddot{C}/2 = \ddot{C}(0)/2.}$$

Let's calculate.
$\theta^b(t)_i = - \tinv \j \a \b \dfs \b \g \a \c
\der \c i.$
$$\eqalign{
\dot{\th}^b(0)_i &=
-\j \a \b \dt(\g \a \c \der \c i)\ds \dft \b
- \half \g \a \c \j \a \b \der \c i \ds \left({d^2f^\b\over dt^2}\right)
\cr
&=
- \j \a \b (\del_\rho \g \a \c )(\j \rho \tau \der \tau j \w a j )
\der \c i \ds \left[ \j \b \mu \der \mu k \ginv k l
(\t a l + s\t b l - \del_l \phi )\right] \cr
&\phantom{= }
- \j \a \b \g \a \c \del_i (\j \c \mu \der \mu j \w a j )
\ds \left[ \j \b \nu \der \nu k \ginv k l
(\t a l + s\t b l - \del_l \phi )\right] \cr
&\phantom{= }
-\half \g \a \c \j \a \b \der \c i \ds \left({d^2f^\b\over dt^2}\right)
\cr
&= - \j \a \b (\del_\rho \g \a \c ) \j \rho \tau \der \tau j
\w a j \der \c i \j \b \nu \der \nu k \ginv k l \t b l \cr
&\phantom{= }
- \j \a \b \g \a \c \j \c \mu (\derr \mu i j \w a j +
\der \mu j \del_i \w a j) \j \b \nu \der \nu k \ginv k l \t b l \cr
&\phantom{= }
-\half \g \a \c \j \a \b \der \c i \ds \left({d^2f^\b\over dt^2}\right).
\cr}$$

\noindent In the above and what follows we use the fact that derivatives
of indices on $M$ with respect to $t$ can be obtained by
$\dft \a \del_\a.$  Also, we will freely use that
$\j \a \c \j \c \b = - \delta^\a {}_\b$ and that $\g \a b$
is hermitian:  $\j \a \mu \g \a \b \j \b \nu = \g \mu \nu.$
Derivatives are finally evaluated at $t = s = 0.$

$$\eqalign{
\dot{\th}^b(0)_i &=
(\del_\rho\g \a \c) \j \rho \tau \der \tau j \w a j
\der \c i \der \a k \w b k
- \g \b \mu \j \b \nu (\derr \mu i j \w a j +
\der \mu j \del_i \w a j) \der \nu k \w b k \cr
&\phantom{= }
-\half \g \a \c \j \a \b \der \c i \ds \left({d^2f^\b\over dt^2}\right)
\cr
&=
(\del_\rho \g \a \c ) \j \rho \tau \der \a k \der \tau j \der \c i
\w b k \w a j \cr
&\phantom{= }
+ \om \nu \mu \der \nu k \der \mu i j \w b k \w a j  \cr
&\phantom{= }
+ \om \nu \mu \der \nu k \der \mu j (\del_i\w a j) \w b k  \cr
&\phantom{= }
-\half \g \a \c \j \a \b \der \c i \ds \left({d^2f^\b\over dt^2}\right)
\cr
}$$

\noindent The third line in the last equality
is zero since $\om \nu \mu \der \nu k \der \mu j$
equals $f^*\omega_{kj},$ which vanishes.

$$\eqalign{
\dot{\th}^b(0)_i &=
 \j \rho \tau (\del_\rho \g \a \c ) \der \a k \der \tau j \der \c i
\w b k \w a j +  \om \nu \mu \der \nu k \der \mu j (\del_i\w a j) \w b k
\cr
&\phantom{= }
+ \half \om \b \c \der \c i \ds \left({d^2f^\b\over dt^2}\right).}$$

At this point we'll need to calculate
$ \ds \left({d^2f^\b\over dt^2}\right) \vert_{t= s= 0}.$
We have
$$\dft \c = \j \c \tau \der \tau l \ginv l m (\t a m +
s \t b m - \del_m \phi ).$$

\noindent Differentiating again, we get
$$\eqalign{{d^2 f^\gamma\over dt^2} &=
\j \c \tau {\del\over \del x^l}\left(\j \tau \nu \der \nu n \ginv n p
(\t a p + s \t b p - \del_p \phi )\right) \ginv l m
\left( \t a m + \t b m - \del_m \phi \right) \cr
& \phantom{= }
+ \j \c \tau \der \tau l \left( -2 h_a{}^{lm} - 2s h_b{}^{lm}\right)
\left( \t a m + s \t b m - \del_m \phi \right) \cr
& \phantom{= }
+ \j \c \tau \der \tau l \ginv l m (- \del_m {d\phi \over dt}).}$$

\noindent Taking the derivative with respect to $s$
gives (note $\phi = 0$ at
t = 0, independent of $s$):
$$\eqalign{
\ds\left({d^2 f^\c \over dt^2}\right)_{t=s=0} &=
\j \c \tau \del_l \left( \j \tau \nu \der \nu n \ginv n p
\t a p\right) \ginv l m \t b m +
\j \c \tau \del_l \left( \j \tau \nu \der \nu n \ginv n p
\t b p\right) \ginv l m \t a m
\cr
& \phantom{= }
-2 \j \c \tau \der \tau l h_a{}^{lm} \t b m
-2 \j \c \tau \der \tau l h_b{}^{lm} \t a m
\cr
& \phantom{= }
- \j \c \tau \der \tau l \ginv l m \del_m \left(
{d^2\phi \over dsdt}\right) \cr
&=
- \del_l(\der \c n \w a n )\w b l - \del_l (\der \c n
\w b n )\w a l \cr
&\phantom{= }
-2 \j \c \tau \der \tau l h_a{}^{lm} \t b m
-2 \j \c \tau \der \tau l h_b{}^{lm} \t a m
\cr
& \phantom{= }
- \j \c \tau \der \tau l \ginv l m \del_m \left(
{d^2\phi \over dsdt}\right).}$$

Let's now plug this result into our formula for $\dot{\th}^b(0)_i.$
$$\eqalign{
\dot{\th}^b(0)_i &=
 \j \rho \tau (\del_\rho \g \a \c ) \der \a k \der \tau j \der \c i
\w b k \w a j +  \om \nu \mu \der \nu k \der \mu j (\del_i\w a j) \w b k
\cr
&\phantom{= }
+ \half \om \b \c \der \c i
\left[- \del_l(\der \c n \w a n )\w b l - \del_l (\der \c n
\w b n )\w a l \right] \cr
&\phantom{= }
+ \half \om \b \c \der \c i
\left[-2 \j \c \tau \der \tau l h_a{}^{lm} \t b m
-2 \j \c \tau \der \tau l h_b{}^{lm} \t a m \right] \cr
&\phantom{= }
+ \half \om \b \c \der \c i
\left[ - \j \c \tau \der \tau l \ginv l m \del_m \left(
{d^2\phi \over dsdt}\right) \right]\cr
&=
 \j \rho \tau (\del_\rho \g \a \c ) \der \a k \der \tau j \der \c i
\w b k \w a j +  \om \nu \mu \der \nu k \der \mu j (\del_i\w a j) \w b k
\cr
&\phantom{= }
-\half \om \b \c \der \c i \left( \derr \b l n \w a n \w b l
+ \derr \b l n \w b n \w a l \right) \cr
&\phantom{= }
- \om \b \c \j \b \tau \der \c i \der \tau l \left( h_a{}^{lm}\t b m
+ h_b{}^{lm}\t a m \right) \cr
&\phantom{= }
- \half \om \b \c \der \c i \j \b \tau \der \tau l \ginv l m
\del_m  \left({d^2\phi \over dsdt}\right) \cr
&=
 \j \rho \tau (\del_\rho \g \a \c ) \der \a k \der \tau j \der \c i
\w b k \w a j +  \om \nu \mu \der \nu k \der \mu j (\del_i\w a j ) \w b k
\cr
&\phantom{= }
-\om \b \c \der \c i \derr \b l n \w a l \w b n \cr
&\phantom{= }
- \g l i  \left( h_a{}^{lm}\t b m
+ h_b{}^{lm}\t a m \right) \cr
&\phantom{= }
- \half \del_i \left({d^2\phi \over dsdt}\right).}$$

Note that $\g l i h_a{}^{lm} \t b m = \g l i h_b{}^{lm} \t a m
= h_{ijk}\w a j \w b k.$ The rest is just algebra in combining the
other terms.  It is trivial to show the following quoted result
in normal coordinates on the
target space (which, since it is K\"ahler, can be chosen simul-
taneously with coordinates adapted to the complex structure),
in which case $h_{ijk} = \om \a \b \der \a i \derr \b j k.$
In any case, it can be shown straightforwardly, in general.
We get, since the
$h$ terms cancel, the following remarkable result.
\eqn\result{{d\t b i\over dt_a}\vert_{t=0} =
- {1\over 2} {\del\over \del x^i} \left({d^2\phi \over dsdt}\right),}
or
\eqn\resultt{\del_a\th^b = d\psi}
at $t=0,$
where $\psi = - \half {d^2\phi \over dsdt}\vert_{t=s=0}.$

\vskip.2in

{\bf Acknowledgements}

We would like to thank M. Bershadsky, D. Morrison, A. Sen, K. Smoczyk,
P. Townsend and C. Vafa for useful discussions.
The research of A.S. is supported in part by DOE grant DOE-91ER40618;
that of S.-T.Y. and E.Z. by grant
DE-F602-88ER-25065. A.S. wishes to thank the physics and mathematics
departments at Harvard, the mathematics department at MIT, and
the physics department at Rutgers for hospitality and support during
the course of this work.

\listrefs
\end